\documentclass[twocolumn,preprintnumbers,prd, nofootinbib]{revtex4}

\usepackage{amssymb,amsmath,graphicx}
\usepackage{epsf}
\usepackage{graphicx,epsfig, epstopdf}
\usepackage{amsfonts}
\usepackage{amssymb}

\def\bk{{\bf k}}

\def\CL{{\cal L}}
\def\CO{{\cal O}}

\newcommand{\nc}{\newcommand}
\nc{\ba}{\begin{eqnarray}}
\nc{\ea}{\end{eqnarray}}
\newcommand\be{\begin{equation}}
\newcommand\ee{\end{equation}}

\newcommand\mPl{{M_{\rm P}}}

\newcommand{\calP}{{\cal P}}
\newcommand{\calR}{{\cal R}}
\nc{\pp}{{\bf{p}}}
\nc{\px}{P_{,X}}
\nc{\pxt}{P_{,X \phi}}
\nc{\pxx}{P_{,XX}}
\nc{\pxxx}{P_{,XXX}}
\nc{\pt}{P_{,\phi}}
\nc{\ptt}{P_{,\phi \phi}}
\nc{\pttt}{P_{,\phi \phi \phi }}
\nc{\pxtt}{P_{,X \phi \phi}}
\nc{\pxxt}{P_{,X X \phi}}

\newcommand{\bea}{\begin{eqnarray}}
\newcommand{\eea}{\end{eqnarray}}
\newcommand{\barr}{\begin{array}}
\newcommand{\earr}{\end{array}}

\begin{document}


\preprint{YITP-13-3, IPM/A-2012/21}

\title{A Single Field Inflation Model with Large Local Non-Gaussianity}

\author{Xingang Chen$^{1}$}
\author{Hassan Firouzjahi$^{2}$}
\author{Mohammad Hossein Namjoo$^{3,4}$}
\author{Misao Sasaki$^{4}$}
\affiliation{$^1$Centre for Theoretical Cosmology, DAMTP, University of Cambridge, Cambridge CB3 0WA, UK}
\affiliation{$^2$School of Astronomy, Institute for Research in
Fundamental Sciences (IPM),
P. O. Box 19395-5531,
Tehran, Iran}
\affiliation{$^3$School of Physics, Institute for Research in
Fundamental Sciences (IPM),
P. O. Box 19395-5531,
Tehran, Iran}
\affiliation{$^4$Yukawa Institute for theoretical Physics,
 Kyoto University, Kyoto 606-8502, Japan}


\begin{abstract}
A detection of large local form non-Gaussianity is considered to be
able to rule out all single field inflation models. This statement is based
on a single field consistency condition. Despite the awareness of some implicit
 assumptions in the derivation of this condition and the demonstration of
corresponding examples that illustrate these caveats, to date there is still
 no explicit and self-consistent model which can serve as a counterexample to
 this statement. We present such a model in this Letter.
\end{abstract}
\maketitle

Primordial non-Gaussianity is an important probe of the inflation models. Different properties of non-Gaussian correlation functions can reveal different physics of the underlying models .
 The non-Gaussianity is usually parameterized by \cite{Chen:2010xka,Liguori:2010hx,Bartolo:2004if}
\ba
\langle \calR_{\bk_1} \calR_{\bk_2} \calR_{\bk_3} \rangle =
\hspace{5.5cm}
\\ \nonumber
(2 \pi)^7 \, f_{NL} \, S(k_1,k_2,k_3) \, \frac{1}{(k_1k_2k_3)^2} \calP_\calR^2
\, \delta^3(\sum_i \bk_i) ~
\ea
where $\calR$ is the curvature perturbation on
comoving slices, $\calP_\calR(k)$ is
its power spectrum per unit logarithmic
momentum interval, and $f_{NL}$ is the amplitude while the function $F(k_1,k_2,k_3)$ is the shape of non-Gaussianity.
In this Letter we concentrate on the so-called local type non-Gaussianity
with the following shape
\ba
S^{\rm loc} = \frac{3}{10} ( \frac{k_1^2}{k_2 k_3} + {\rm 2~perm.}) ~,
\ea
where ``perm" stands for the cyclic permutation of three momenta. This shape peaks at the squeezed limit.
In contrast, other shapes that we will mention later include the equilateral-like shape
\begin{align}
S^{\rm equil} =
-6 ( \frac{k_1^2}{k_2 k_3} + {\rm 2~ perm.})
+ 6 (\frac{k_1}{k_2} + {\rm 5~ perm.} ) -12 ~,
\end{align}
which peaks at the equilateral limit and typically arises in non-slow-roll models with non-canonical kinetic term; and the folded-shape
\begin{align}
S^{\rm fold} = 6 ( \frac{k_1^2}{k_2k_3} + {\rm 2~perm.}) -6 (\frac{k_1}{k_2} + {\rm 5~perm.}) +18 ~,
\end{align}
which peaks at the folded limit ($k_1+k_2=k_3$, and cyclic) and qualitatively describes certain features in models with non-Bunch-Davies vacuum.

There is an important statement on how the local non-Gaussianity can be used to distinguish inflation models:

0) {\em A detection of a large local non-Gaussian component in the bispectrum can rule out all single field inflation models} \cite{Creminelli:2004yq,Komatsu:2009kd}.

The size of local non-Gaussianity which can be detected with high confidence level in the near future is $f^{\rm loc}_{NL} \gg 1$. By single field inflation models, we include not only the slow-roll single field models with Bunch-Davies (BD) vacuum, but also all other inflation models that have one field responsible for both the inflation and creation of curvature perturbation. The statement $0)$ is based on Maldacena's consistency condition for the single field models [\citenum{Maldacena:2002vr}, \citenum{Creminelli:2004yq}],
\ba
\label{consistency_cond}
&&
\langle \calR_{\bk_1} \calR_{\bk_2} \calR_{\bk_3} \rangle
\cr
&&\quad
\to
(1-n_s) \frac{(2\pi)^7}{4k_1^3 k_3^3} \calP_\calR(k_1) \calP_\calR(k_3)
\delta^3(\sum_i \bk_i) ~,
\ea
in which $k_3 \ll k_1=k_2$ and $n_s$ is the spectral index.
Since the momentum-dependence on the RHS of (\ref{consistency_cond}),
$\sim 1/k_1^3k_3^3$, takes the scale-invariant local form, the size of the local non-Gaussianity in the squeezed limit ($k_3 \ll k_1=k_2$) is $f^{\rm loc}_{NL} \sim 1-n_s$, which is of order the slow-variation parameter $\CO(\epsilon)\sim \CO(0.01)$ at the leading and non-oscillatory order. Therefore these models predict very small local non-Gaussianity.

The derivation of this condition relies on a very general assumption: for single field, the only effect of a long wavelength mode on short wavelength modes is to provide a constant rescaling of the background scale factor. Nonetheless, despite of its generality, it has been noticed that there are a couple of implicit assumptions underlying the derivation of this condition \cite{Chen:2010xka,Namjoo:2012aa,NGworkshop2012}:

A) {\em There is no large correlation between modes when all modes are sub-horizon. So we only need to consider starting from when the long wavelength mode is outside the horizon} (Sec.~9.2 in \cite{Chen:2010xka});

B) {\em The amplitude of the superhorizon mode remains constant, namely, the attractor solution is reached} \cite{Creminelli:2004yq,Namjoo:2012aa}.

Although it may be highly unnatural that these assumptions are broken for infinitely squeezed configuration, or in the whole 60 $e$-folds range of inflation, one can consider to invalidate them in a smaller range so that the condition (\ref{consistency_cond}) is violated up to some finitely-squeezed limit. Such a violation is relevant for realistic experiments which can only measure finitely-squeezed limits.

So we emphasize that the statement 0) and the condition (\ref{consistency_cond}) are not equivalent. A simple fix is to supplement the extra conditions A) and B) to the statement 0). But more interestingly it has become an important question whether we can find counterexamples to 0) by exploring the caveats A) and B).
Despite of many efforts, to date no explicit and self-consistent example is known that can invalidate the statement 0), although there {\em do exist} many examples that violate the consistency condition (\ref{consistency_cond}) through violating the assumptions A) and/or B). The following is a brief summary of this situation.

$\hat {\rm A}$) For example, subhorizon modes can have large non-Gaussian
correlations if the vacuum is non-BD and has a negative energy
component \cite{Chen:2006nt,Holman:2007na,Meerburg:2009ys,Chen:2010bka,Agullo:2010ws,Ganc:2011dy,Chialva:2011hc,Ganc:2012ae, Agullo:2012cs,Agarwal:2012mq, Ashoorioon:2010xg}.
An observational signature is enhanced 3pt correlation in folded
triangle configurations, which include the squeezed configuration so could have
overlap with the local form. However the non-BD vacuum component
in \cite{Chen:2006nt,Holman:2007na,Meerburg:2009ys,Agullo:2010ws,Ganc:2011dy,Chialva:2011hc,Ganc:2012ae,Agullo:2012cs,Agarwal:2012mq} is put in by hand and cut off at a specific time
because it cannot be extended to infinite past. This procedure illustrates certain qualitative features of this type of non-Gaussianity, but is not fully self-consistent in relevant details.
Around the cutoff, the full set of
equations of motion are not solved.
It is the assumption of this cutoff procedure that the non-Gaussian correlation
is integrated starting from the cutoff, but before and especially around the cutoff,
the physics may be already important if done consistently and change the final non-Gaussian profile. Indeed, in a fully consistent toy example studied
in \cite{Chen:2010bka}, the non-Gaussianity still peaks at the folded limit,
but at the same time acquires an overall  scale-dependent oscillation around zero.
Therefore a scale-invariant local template is not able to pick up such signals
even though they violate the consistency condition at some finitely-squeezed
configurations. What should happen to other possible consistent examples
remains an open question.

For another example, subhorizon correlation is also possible
from feature models
such as the resonant model \cite{Chen:2008wn,Flauger:2010ja}
 or sharp feature model \cite{Chen:2006xjb,Adshead:2011jq,Arroja:2011yu,Arroja:2012ae},
 which can violate the condition (\ref{consistency_cond}) for
finitely-squeezed configurations from the leading BD component.
In these examples,  a scale-dependent oscillation around zero also exists.

$\hat {\rm B}$) The possibility of non-attractor solutions has also been explored. There are examples such that a transient stage to inflation leads to non-Gaussianities with  scale-dependent oscillations around zero \cite{Chen:2008wn}, or examples that generate non-oscillatory (yet still strongly scale-dependent) local non-Gaussianity but only in non-inflationary and contracting background \cite{Khoury:2008wj,Cai:2009fn}. Recently, a non-attractor inflation model with scale-invariant power spectrum and bispectrum is presented \cite{Namjoo:2012aa}, see also \cite{Kinney:2005vj,Martin:2012pe},
which clearly demonstrates the compatibility between the scale-invariance and the caveat B). But the model only produces a relatively small non-Gaussianity, $f^{\rm loc}_{NL} = 2.5$. Hence the statement 0) still holds.

Therefore a continued search in both possibilities, or an investigation for other possibilities \cite{footnoteOther}, become very important. Especially if such a local component were discovered in future observations, these studies would help us to understand its precise implications.

In the rest of this Letter, we present a self-consistent counterexample to the statement 0) by exploring the second caveat B) along the line of Ref.~\cite{Namjoo:2012aa}.
We emphasize that, for the purpose of this Letter we allow various fine-tunings to get the right amount of $e$-folds and the scale-invariant power spectrum, and we do not address the UV completion aspects of the model, because these are {\em not the concerns} of either the statement 0) and the condition (\ref{consistency_cond}), or the assumptions A) and B). Possible improvement on such issues can be interesting subjects for future work.

We consider general single field inflation with non-canonical kinetic term [\citenum{Garriga:1999vw}, \citenum{Chen:2006nt}]. For reasons that will become clear, we seek a model in which the sound speed $c_s\ll 1$ is constant but the slow-roll parameter
$\epsilon = - \dot H/H^2$ decays rapidly with a constant rate, $\eta = \dot \epsilon/H \epsilon \sim \mathrm{const}$. The Lagrangian of our model is
\ba
\label{model}
P= X+ \dfrac{ X^\alpha }{M^{4\alpha - 4}}  -V(\phi) \,  , \, V(\phi)= V_0 + v \left( \dfrac{ \phi }{M_P} \right)^\beta ,
\ea
in which $X = -\frac{1}{2} \partial_\mu \phi \partial^\mu \phi $ and $M, \alpha, v, V_0$ and $\beta$ are free constant parameters. We choose initial conditions such that the constant term in the potential dominates the total energy density
 and drives the inflation. We assume that in a non-attractor transient inflationary phase, the canonical kinetic term $X$ is sub-dominant.
We check the consistency of this assumption below.
After sufficient decay of the kinetic energy, the linear term dominates which gives a second, conventional slow-roll inflationary phase.
Since the superhorizon modes are frozen during the second phase, taking care of their evolution in the first non-attractor phase is enough. The background equations of motion are
\bea
&&
3 \mPl^2 H^2 = 2X P_{,X}-P ~, \quad \mPl^2 \dot H = - X P_{,X} ~,
\\
&&
{P_{,X}}c_s^{-2}  \dot X + 6 H  X P_{,X}- P_{,\phi} \dot \phi =0 ~.
\label{KG}
\ea
The sound speed is
$c_s^2 = {\px}/{(\px+2 X \pxx)}$,
and for future references,
$\Sigma = X P_{,X} + 2 X^2 P_{,XX}=H^2\epsilon/c_s^2$,
$\lambda = X^2 P_{,XX} + \frac{2}{3} X^3 P_{,XXX} $.
In the non-attractor phase where the non-linear kinetic term dominates, we have
\bea
c_s^2 \simeq {1}/{(2\alpha - 1)} ~.
\eea
For $\alpha \gg 1$, we have a constant $c_s \ll 1$ as desired. The expression for $\lambda$ also simplifies to
$\lambda/\Sigma = (1-c_s^2)/(6 c_s^2)$.

We solve the background equations by proposing the ansatz,
$\phi = \phi_0 a^\kappa$, in which $\kappa$ is a constant. Physically this corresponds to a fine-tuning of the inflationary potential
and the background initial conditions.
Plugging this ansatz in (\ref{KG}) and
noting that $H$ is nearly constant,
we constrain some parameters in the potential,
\ba \label{sol2}
\beta = 2 \alpha~, \quad
v =  - \dfrac{M^4}{c_s^2} \left( \dfrac{V_0 \kappa^2}{6 M^4} \right)^\alpha (1+{3 c_s^2}{\kappa}^{-1} ) ~.
\ea
Besides that we have $
\epsilon = {XP_{,X}}/{\mPl^2 H^2} \propto a^{2 \alpha \kappa}$.
So the constant $\kappa$ also yields a constant $\eta$,
\bea
\eta=2\alpha\kappa
\eea
as we desired.
As we will see later,
the scale-invariance of the power spectrum requires $\eta\simeq -6$.

Using these results one can easily check that
${X^\alpha/M^{4 \alpha-4}} \simeq - c_s^2 (1+{3c_s^2}{\kappa}^{-1} )^{-1}
{v \left(\phi/\mPl \right)^\beta} \ll -v \left(\phi/\mPl \right)^\beta \ll V_0$.
That is, the total energy density is dominated by the potential.

In the above, we assumed the canonical kinetic term is negligible,
$X \ll X^\alpha/M^{4\alpha -4}$, or $X/M^4>1$ for $\alpha\gg1$.
 Using the above ansatz this condition translates into
$  \phi \sqrt{{V_0 \kappa^2}/{6\mPl^2 M^4}}  > 1 $ .
This condition breaks down at $\phi_*$ given  by
$\phi_*/\mPl \sim M^2 \sqrt{{6}/{V_0 \kappa^2}} $.
After $\phi_*$, we can have a slow-roll phase, by suitable initial conditions. If this does not happen, we lose our analytic control on the full solution.

\begin{figure}[t]
\includegraphics[ scale=.315]{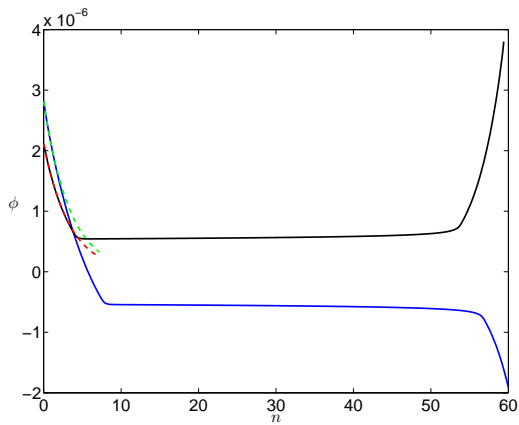}
\caption{Inflaton $\phi$ versus inflationary $e$-fold $n$. The upper solid black curve
and the lower solid blue curve represent  the undershoot and overshoot case, respectively. The dashed lines are the analytic ansatz for the non-attractor phase. Parameters are $V_0 \simeq 6 \times 10^{-4}$, $M=5 \times 10^{-5}$, $\alpha = 10$ and $\eta = -6$, ($M_P \equiv 1$).}
\label{phin}
\end{figure}

\begin{figure}[t]
\includegraphics[ scale=.3 ]{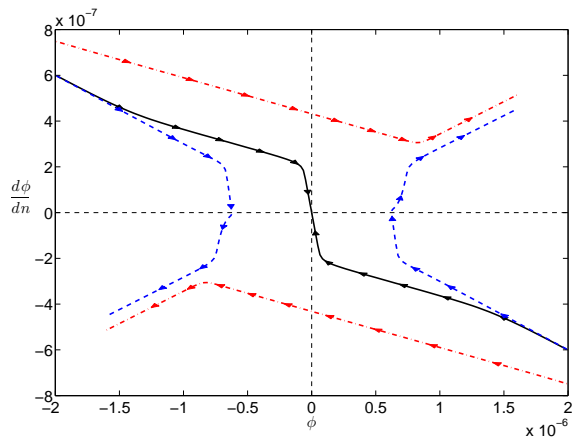}
\caption{Phase diagram for the same model as in Fig.~\ref{phin}.
Dashed blue and dashed-dotted red curves represent the undershoot and overshoot case, respectively. The black solid curve is the separator of different types of trajectories. }
\label{phase}
\end{figure}

The picture we present here is that the inflaton rolls up the potential
at the first non-attractor inflation phase.
This is in contrast to the usual picture that the inflaton rolls down the potential
in the attractor phase.  Depending on initial conditions the inflaton can
have different trajectories. One possibility is that it stops
somewhere before crossing the origin, and then starts to roll down on
the same side of the potential (i.e.~undershoot). Another possibility
is to cross the top of the potential and then roll down on the other
side (i.e.~overshoot). The non-attractor inflationary phase when the inflaton
rolls up the potential is the one we concentrate in this study.
This picture is numerically simulated in Figs.~\ref{phin} and \ref{phase}.

Now, let us look at the perturbation.
 The quadratic perturbative Lagrangian for the comoving
curvature perturbation is \cite{Garriga:1999vw}
\bea
\CL_2 = a^3 \frac{\epsilon}{c_s^2} \dot\calR^2 - a\epsilon (\partial\calR)^2 ~.
\eea
Using $\epsilon \propto a^\eta$ and assuming the Bunch-Davies vacuum deep inside the horizon,
the mode function is given by
\bea
\label{mode1}
\calR_k = C_k x^\nu H_\nu^{(1)}(x) ~,
\eea
where we defined
$x\equiv - c_s k \tau$, $\nu \equiv {(3+\eta)}/{2}$,
and
$\vert C_k \vert^2 = k^{-2\nu} \pi c_s^{2-2\nu} H^2 (-\tau_i)^{3-2\nu}/(8\epsilon_i M_P^2)$.
Here the subscript $i$ indicates some initial time during the non-attractor phase.
Expanding the Hankel function with rank $\nu <0$ for small $x$, one obtains the power spectrum in terms of the physical variables at the end of the non-attractor phase (indicated by the subscript $e$),
\ba
{\calP_R(k)}\simeq
\dfrac{\Gamma(\vert \nu \vert)^2 }{\pi^3 2^{2\nu+4} } \left(\dfrac{H_e}{\mPl} \right)^2 \dfrac{1}{c_s\epsilon_e} \left( \dfrac{c_sk}{H_e a_e} \right)^{3+2\nu} ~.
\ea
Note that during the non-attractor phase the curvature perturbation grows very rapidly on super-sound-horizon scales due to the fast decay of $\epsilon$.
This phase terminates once the slow-roll phase starts and the curvature perturbation on super horizon scales freezes afterwards.

The spectral index is given by
$n_s-1 \simeq 3+2\nu = 6+\eta$.
Very interestingly, we simply need $\eta \simeq -6$ to
have a near scale invariant power spectrum.
In the following we focus on this case with its simplified mode function
\ba
\label{mode2}
\calR_k = C_k \sqrt{\dfrac{2}{\pi}} \,
 \dfrac{e^{-i c_s k \tau}}{(-c_s k \tau)^3} (-1-i c_s k \tau) ~.
\ea

It is also instructive to consider the reason for the scale-invariance in terms
of the scalar field perturbation in the spatially flat gauge, where the scalar fluctuation is defined in terms of $\phi=\phi_0 +\delta\phi$.
At the leading order, $\calR = -H \delta\phi/\dot\phi_0$.
For both the conventional attractor
 slow-roll case and the previous non-attractor
example \cite{Namjoo:2012aa}, the scale-invariance
comes from the fact that $\delta\phi$ is approximately constant on
superhorizon scales.
In our model, the background $H$ and $\dot\phi$ are slowly varying, while
$\delta\phi$ grows as $a^3$ both at sub-sound-horizon and super-sound-horizon scales relative to the conventional case.
Alternatively however, if one rewrites everything by introducing a redefined scalar
field $\sigma\propto \phi^\alpha$, one finds that $\delta\sigma$
behaves like a canonical scalar field with the effective mass-squared
$m_\sigma^2=H^2\eta(\eta+6)/4$, provided that the background
evolves as $\sigma \propto a^{\eta/2}$. The only difference from
the canonical case is the sound velocity.
Thus for $\eta\simeq-6$, both the background $\sigma$
and the perturbation $\delta\sigma$ behave like those in the
non-attractor model discussed in \cite{Namjoo:2012aa},
and the scale-invariance is recovered.
This analysis also suggests that the background trajectory is fine-tuned as noted.
As in the previous case \cite{Namjoo:2012aa}, because of the relation
$\calR=-H\delta\sigma/\dot\sigma\propto a^{-\eta/2}\delta\sigma$, the amplitude
of $\delta\sigma$ at the sound-horizon crossing must be small enough.
This condition can be satisfied by choosing a low scale $H$,
resulting in a negligible tensor-to-scalar ratio.

The cubic perturbative Lagrangian for general single field inflation is
\cite{Seery:2005wm,Chen:2006nt}
\begin{eqnarray} \label{action3}
{\cal L}_3&=&
-a^3 (\Sigma(1-\frac{1}{c_s^2})+2\lambda)\frac{\dot\calR_n^3}{H^3}
+\frac{a^3\epsilon}{c_s^4}(-3+3c_s^2)\calR_n\dot\calR_n^2
\nonumber \\ &+&
\frac{a\epsilon}{c_s^2}(1-c_s^2)\calR_n(\partial\calR_n)^2
+ \dots ~,
\end{eqnarray}
with the field redefinition,
\bea
\calR=\calR_n+\frac{\eta}{4c_s^2}\calR_n^2+\frac{1}{c_s^2H}\calR_n\dot{\calR_n}
+ \dots ~.
\label{redefinition}
\eea
Terms not listed in Eqs. (\ref{action3}) and (\ref{redefinition})
are suppressed by either $\CO(\epsilon^2)$ or spatial derivatives.

If non-Gaussianity is sourced by large interaction terms in the inflaton sector, the interaction introduced by gravity becomes negligible. It is often convenient to go to the spatially flat gauge, where one can simply perturb $P(X,\phi)$ with respect to $\delta\phi$ and get
\bea
\CL_3 = a^3 \frac{2\lambda}{\dot\phi_0^3} \delta{\dot\phi}^3
- a \frac{\Sigma (1-c_s^2)}{\dot\phi_0^3} \delta{\dot\phi} (\partial \delta\phi)^2 ~,
\label{action_deltaphi}
\eea
where the derivatives of $P$ with respect to $\phi$ is ignored since they are not enhanced by $c_s^{-2}$. Integrating by part one can check that, for $H,\dot\phi, \eta, c_s \sim {\rm const.}$ and $\lambda/\Sigma, c_s^{-2} \gg 1$,  the above two descriptions (terms listed in spatially flat gauge vs. comoving
gauge) are equivalent at the leading order.

The two terms in the field redefinition give local shape non-Gaussianity as usual.
In the familiar attractor single field case \cite{Chen:2006nt}, the terms in (\ref{action3}) or (\ref{action_deltaphi}) are responsible for two independent leading order equilateral-like shapes.
Interestingly, the physical consequence of these terms becomes very different here. The last term in either (\ref{action3}) or (\ref{action_deltaphi}) is now subdominant due to the spatial derivative.
The rest of the terms still contribute to large non-Gaussianity but the shape is no longer equilateral.
In \cite{Chen:2006nt}, the super-sound-horizon mode is constant and
cannot generate large correlations with sub-sound-horizon modes; so
a large bispectrum comes from when all three modes exit the sound horizon
at about the same time, resulting in equilateral shapes.
In contrast, here both the modes (\ref{mode2}) and their correlations are
still growing after sound-horizon exit, so we expect the non-Gaussianity
to carry a significant local component.
Detailed calculation (see \cite{Chen:2010xka} for method) reveals that
all these leading shapes are exactly of the local form, and scale invariant.
For $\eta \simeq -6$ and $c_s^2 \ll 1$, the final result, summing over the four terms in the comoving gauge or equivalently only one term in the spatially flat gauge, is
\ba
f_{NL}^{\rm loc} \simeq \dfrac{5}{4 c_s^2} ~.
\ea
This is our main result. The size of the {\em local} non-Gaussianity is
enhanced by $1/c_s^2$ which for sufficiently small $c_s$ can be detectable
in near future observations. As a result, a detection of large local
non-Gaussianity alone will not rule out all single field inflation models.\\


{\bf Acknowledgment} We would like to thank A. A. Abolhasani,
M. Noorbala and M. Yamaguchi for discussions.
X.C, H.F. and M.H.N. would like to thank the Max Planck
Institute for Astrophysics and the Yukawa Institute for Theoretical Physics at
Kyoto University for their hospitality where this work was in progress during
 the workshop ``Critical tests of inflation using non-Gaussianity" and
``Gravity and Cosmology 2012'', No. YITP-T-12-03.
X.C. is a Stephen Hawking Advanced Fellow.
This work was also supported in part by JSPS Grant-in-Aid for Scientific
Research (A) No.~21244033.


\end{document}